\documentclass[apjl,tighten]{emulateapj}

\usepackage{newtxtext,newtxmath}

\usepackage[T1]{fontenc}
\usepackage{ae,aecompl}
\usepackage{color}
\usepackage[normalem]{ulem}
\usepackage[breaklinks,colorlinks,urlcolor=blue,citecolor=blue,linkcolor=blue]{hyperref}

\usepackage{graphicx}	
\usepackage{amsmath}	
\usepackage{amssymb}	

\newcommand{\Msun}{\ensuremath{\mathrm{M}_\odot}}
\newcommand{\ud}{\ensuremath{\mathrm{d}}}

\begin{document}

\title[Black hole formation and fallback]{Black hole formation and fallback during the supernova explosion of a $40 \,\Msun$ star}

\author{
Conrad Chan\altaffilmark{1,2},
Bernhard M\"uller\altaffilmark{1,3},
Alexander Heger\altaffilmark{1,4,5},
R\"udiger Pakmor\altaffilmark{2}
and Volker Springel\altaffilmark{2,6,7}}

\affil{
\altaffilmark{1}{Monash Centre for Astrophysics, School of Physics and Astronomy, Monash University, Victoria 3800, Australia;
\href{conrad.chan@monash.edu}{conrad.chan@monash.edu}
}
\\
\altaffilmark{2}{Heidelberger Institut f\"ur Theoretische Studien, Schloss-Wolfsbrunnenweg 35, 69118 Heidelberg, Germany
}
\\
\altaffilmark{3}{Astrophysics Research Centre, School of Mathematics and Physics, Queen's University Belfast, Belfast, BT7 1NN, United Kingdom}
\\
\altaffilmark{4}{School of Physics \& Astronomy, University of Minnesota, Minneapolis, MN 55455, U.S.A.}
\\
\altaffilmark{5}{Center for Nuclear Astrophysics, Department of Physics and Astronomy, Shanghai Jiao-Tong University, Shanghai 200240, P. R. China.}
\\
\altaffilmark{6}{Zentrum f\"ur Astronomie der Universit\"at Heidelberg, Astronomisches Recheninstitut, M\"onchhofstr. 12-14, 69120 Heidelberg, Germany}
\\
\altaffilmark{7}{Max-Planck-Institut f\"ur Astrophysik, Karl-Schwarzschild-Str. 1, 85740 Garching bei M\"unchen, Germany}
}

\date{Accepted XXX. Received YYY; in original form ZZZ}


\label{firstpage}

\begin{abstract}
Fallback in core-collapse supernovae is considered a major ingredient for explaining abundance anomalies in metal-poor stars and the natal kicks and spins of black holes (BHs).  We present a first 3D simulation of BH formation and fallback in an ``aborted'' neutrino-driven explosion of a $40$ solar mass zero-metallicity progenitor from collapse to shock breakout.  We follow the phase up to BH formation using the relativistic \textsc{CoCoNuT-FMT} code.  For the  subsequent evolution to shock breakout we apply the moving-mesh code \textsc{Arepo} to core-collapse supernovae for the first time.  Our simulation shows that despite early BH  formation, neutrino-heated bubbles can survive  for tens of seconds before being accreted,  leaving them sufficient time to transfer part of  their energy to sustain the shock wave as is  propagates through the envelope.  Although the  initial net energy ($\sim 2$ Bethe) of the neutrino-heated ejecta barely equals the binding energy of the envelope, $11\,\Msun$ of hydrogen are still expelled  with an energy of $0.23$ Bethe.  We find no significant mixing and only a modest BH kick and spin, but speculate that stronger effects could occur for slightly more energetic explosions or progenitors with less  tightly bound envelopes.
\end{abstract}


\maketitle



\section{Introduction}
 Often the collapse of massive stars results in the formation of a black hole (BH) instead of a neutron star.  Observations of supernova progenitors suggest that a large fraction of stars above $15\,\Msun\ldots 18\,\Msun$ collapse quietly to BHs \citep{Smartt2015}, and first direct evidence for the disappearance of red giants is currently being discussed \citep{Adams2017}.

It appears, however, that BH formation does not always proceed quietly without any mass ejection.  Even if no neutrino-driven or magnetohydrodynamically-driven engine explodes the star, the loss of gravitational mass due to neutrino emission may lead to the (partial) ejection of the hydrogen envelope, giving rise to a faint red transient \citep{Nadyozhin1980,Lovegrove2013}.

BH formation may also occur in a full-blown supernova explosion.  In the collapsar model \citep{MacFadyen1999} for gamma-ray bursts and hypernovae, the formation of a BH with an accretion disk is central to the mechanism itself.  In neutrino-driven explosions or magnetorotational explosions involving rapidly spinning neutron stars, BH formation can occur due to \emph{fallback}, which pushes the neutron star above its maximum mass limit.  Fallback can be ``prompt'' \citep{Wong2014} when accretion onto the neutron star remains strong in the first seconds of an incipient explosion, or occur at late times times when the reverse shock decelerates outgoing shells below the escape velocity \citep{Chevalier1989}. Particularly strong fallback is expected for weak explosions with energies comparable to the binding energy of the stellar envelope \citep{Zhang2008}. 

There is a body of evidence for fallback in supernova explosions.  Transients with unusually low explosion energies and nickel masses have been interpreted as being due to fallback \citep{Zampieri+2003,Moriya2010}. Further evidence comes from extremely metal-poor (EMP) stars that were presumably polluted by the ejecta of just one or at most a few supernovae from Population~III (Pop~III) progenitors.  Abundance anomalies in such EMP stars like high [C/Fe] and [O/Fe] ratios have been interpreted as resulting from the removal of most of the iron-group ejecta from the polluting supernova by fallback but allowing for mixing from some inner shells into the ejecta (mixing-fallback model; \citealp{Nomoto+2006}).  The star J031300.36-670839.3 is the most famous example.  To date, no iron at all was detected, only a very low upper limit for iron could be determined.  It is thought to originate from a fallback supernova from a  $60\,\Msun$ \citep{Keller2014} or $40\,\Msun$ \citep{Bessell+2015} star.  Even more direct evidence for BH formation associated with a supernova comes from the pollution of companions in some X-ray binaries \citep{Israelian1999}.

If there is a BH formation channel via fallback, then this also has implications for our understanding of the BH mass and spin distributions, which are now becoming accessible through GW detections \citep{Abbott2016} in addition to measurements of BH properties within the Milky Way (masses: \citealp{Oezel2010}, spins: \citealp{Liu2008,MM2015}).  It has also been suggested that asymmetric fallback can give rise to large BH kicks \citep{Janka2013NatalSupernovae}, which would have repercussions on the evolution of binary systems with fallback supernova progenitors.

Understanding BH formation, fallback, and mixing in core-collapse supernovae calls for detailed numerical simulations.  Late fallback due to outflow deceleration by the reverse shock can be studied \citep{Zhang2008,Ertl2016b} in spherical symmetry (1D), but multi-dimensional (multi-D) models are required to address mixing and the effect of the fallback on BH kicks and spins.  Mixing and prompt fallback are intimately connected to the asymmetries imprinted on the explosion by the supernova engine (whether neutrino-driven or MHD-driven).

Although multi-D  simulations of fallback and mixing due to the Rayleigh-Taylor instability -- typically with a focus on massive Pop~III progenitors -- were conducted in 2D  \citep{Joggerst2009,Chen2017} and 3D \citep{JAW2010,Joggerst2010b}, these were based on artificial 1D explosion models for the first seconds of the supernova.  These models thus miss the large-scale seed asymmetries that are provided by the supernova engine, and that have been shown to decisively influence mixing out to late times in simulations starting from more consistent multi-D neutrino-driven explosions models of SN~1987A \citep{WMJ2015} and Cas~A \citep{Wongwathanarat2017}.  They must also make \emph{ad hoc} assumptions about the energetics of the explosion to achieve the desired fallback.

Here we perform the first 3D simulations of a fallback supernova from the collapse stage through the onset of a neutrino-driven explosion, further through BH formation, and continued until shock breakout.  Motivated by the abundance fits of \citet{Keller2014} and \citet{Bessell+2015} for the polluter of J031300.36-670839.3, we consider a $40 \Msun$ Pop~III progenitor, although we do not attempt to reproduce any particular abundance pattern in this paper.  We rather seek to develop a basic understanding of explosions with strong fallback in multi-D: How is the energy transferred from the engine to the envelope?  Are the $\mathord{\sim} 10^{51}\,\mathrm{erg}$ from a neutrino-driven engine sufficient to unbind even part of an envelope with a much higher binding energy?  If so, can efficient mixing develop during the explosion? What kick and spin is imparted onto the BH by asymmetric fallback?

\section{Numerical Methods}

\subsection{Collapse to Black Hole Formation}

We model the collapse of a non-rotating metal-free $40\,\Msun$ progenitor evolved with the stellar evolution code \textsc{Kepler} \citep{HW2010}.  At the onset of  core collapse, the model is mapped into the relativistic neutrino hydrodynamics code \textsc{CoCoNuT-FMT} \citep{Mueller2015a}.  The model has been simulated on a spherical polar  grid with a resolution of $N_\mathrm{r} \times N_\theta \times N_\varphi = 550 \times 128 \times 256$ zones out to a radius of $10^5\,\mathrm{km}$ using a mesh coarsening scheme near the polar axis. 

At high densities, we employ the nuclear equation of state (EoS) of \citet{Lattimer1991} with a bulk incompressibility modulus of $K=220 \, \mathrm{MeV}$, which allows for a maximum baryonic neutron star mass of  $2.44\,\Msun$.  To precipitate shock revival, we artificially increase the strangeness contribution to the axial vector coupling for neutral current neutrino-nucleon scattering to an (unphysically large) value of $g_\mathrm{A,s}=-0.2$ as in \citet{Melson+2015}.  This is justified as long as we merely seek to provide plausible initial conditions for the simulation of fallback and mixing after BH formation \emph{assuming} that a neutrino-driven explosion occurs. 

\subsection{Black Hole Formation to Shock Breakout}

After BH formation, we follow the evolution of the explosion until shock breakout using the quasi-Lagrangian moving-mesh hydrodynamics code \textsc{Arepo} \citep{Springel2010}.   \textsc{Arepo} solves the  fluid equations in a finite-volume approach on an  unstructured Voronoi mesh using adaptive time-stepping, second-order spatial reconstruction, and the HLLC Riemann solver.  The position of the grid cells track the flow, providing resolution where it is needed in order to follow any mixing that develops in the frame of the rapidly expanding shock.  The unstructured mesh of \textsc{Arepo} has no preferred directions, eliminating grid-alignment artifacts that sometimes plague traditional Eulerian hydrodynamics codes. We use the Helmholtz EoS \citep{TS2000}, which includes degeneracy and radiation pressure effects.

We account for the gravity of the BH by including the monopole potential of a point mass, and the self-gravity between gas cells is solved following \citet{Springel2010} using a tree-based algorithm with a  variable softening length with a spline function that varies from $10^3\,\mathrm{km}$ to $10^{8}\,\mathrm{km}$.

Except for the gravitational potential, periodic boundaries are used outside a box of $5\times10^{7}\,\mathrm{km}$ width, which contains the entire star.  An exponential isothermal atmosphere is adopted outside the star. 

In order to map the \text{CoCoNuT} model into \textsc{Arepo}, we initialize a mesh generating point at the geometric center of each grid cell.  To achieve the target resolution, we apply cell refinement criteria based on the desired particle mass, such that the resolution is concentrated at regions of higher density.  Since the \textsc{CoCoNuT} model only includes the region inside a radius of  $R=10^{5}\,\mathrm{km}$, we add the original 1D pre-supernova profile outside, which smoothly fits the 3D models since no significant changes have occurred at the fitting radius at the time of mapping.  In the outer layers, we adopt a mesh-generating point configuration similar to that introduced by \cite{ORP2015}, using nested shells of HEALPix configuration \citep{Gorski+2004}.  Due to small differences in the low-density EoS, we map the temperature instead of the internal energy density, which makes the mapping slightly non-conservative, but small and acceptable conservation errors are unavoidable anyway when mapping a GR model to a Newtonian code: We find a $0.3\,\%$ change in total mass of $0.044\,\Msun$, and  $4.5\,\%$ change in diagnostic explosion energy (\S~\ref{sec:results}) of $0.095\,\textrm{B}$.

\begin{figure}
    \includegraphics[width=\columnwidth]{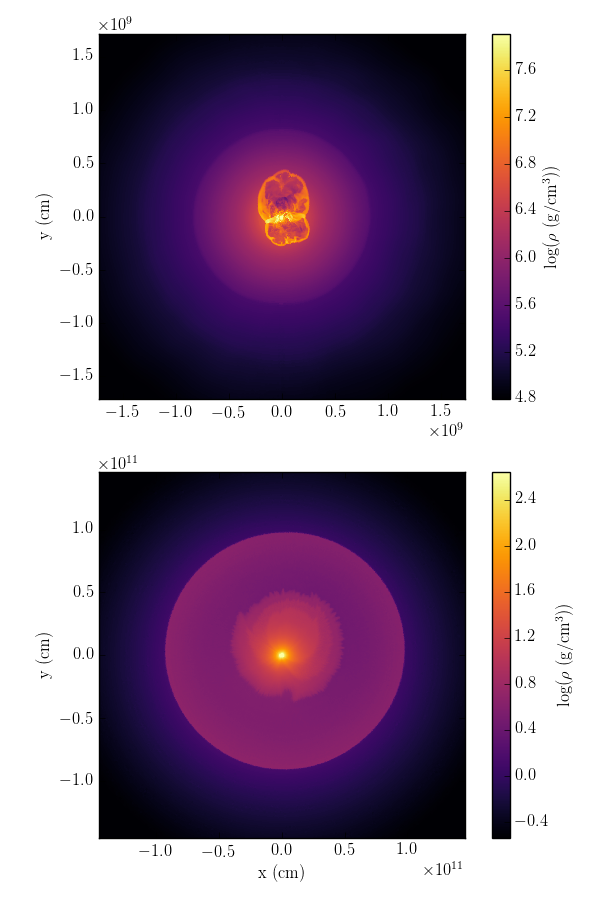}
    \caption{\textsl{Top:} Density slice through the origin at BH formation (initial conditions for \textsc{Arepo} run).  \textsl{Bottom:} Density slice at $215\,\mathrm{s}$, shortly after the  explosion energy converges.  At this stage, the shock is roughly spherically symmetric. \label{fig:initial}}
\end{figure}

To model the accretion onto the BH, we continuously remove all mass below a radius $R_\textrm{acc}$ and add it to the central point mass. For testing purposes, we investigated the effect of varying the infall radius from $3\,R_\textrm{s}$ to $30\,R_\textrm{s}$, where $R_\textrm{s}$ is the Schwarzschild radius of the BH. Due to the highly supersonic infall velocities in the central regions, the evolution of the stellar envelope is insensitive to the infall radius.

\section{Results}
\label{sec:results}

Due to the high progenitor mass, the model maintains high accretion rates after core bounce, and the proto-neutron star quickly grows in mass, reaching a baryonic mass of $2\,\Msun$ in less than $0.2\,\mathrm{s}$, and contracts rapidly.  Such conditions are conducive \citep{Mueller2012} to violent activity of the standing accretion shock instability (SASI; \citealp{Blondin2003}).  This helps stabilize the retraction of the average shock radius at about $100\, \mathrm{km}$ around $0.2\,\mathrm{s}$ after bounce.  High electron flavor neutrino luminosities $\mathord{\gtrsim} 10^{53}\,\mathrm{erg} \,\mathrm{s}^{-1}$, hard neutrino spectra, a small gain radius, and stabilization of the shock radius by the SASI result in high heating efficiencies and eventually in shock revival $0.35\,\mathrm{s}$ after bounce.  Due to our artificial reduction of the neutrino-nucleon scattering cross section and due to the approximations in the \textsc{FMT} transport scheme, more rigorous simulations are needed to test whether this scenario for shock revival for rapidly accreting proto-neutron stars is indeed viable. It is, however, not implausible. The trend towards powerful SASI activity and high heating efficiencies for massive neutron stars is in line with other simulations, especially with more systematic 2D studies \citep{Summa2016,Burrows2016} that often obtain explosions more readily for more massive progenitors.  Even if our neutrino treatment overestimates the heating conditions, it is plausible that other physical ingredients could drive our model into a neutrino-driven runaway, such as rapid rotation \citep{Summa2017}, auxiliary magnetic field effects \citep{Obergaulinger2014}, or enhanced neutrino luminosities due to novel microphysics at high densities \citep{Bollig2017}. 

\begin{figure}
	\includegraphics[width=\columnwidth]{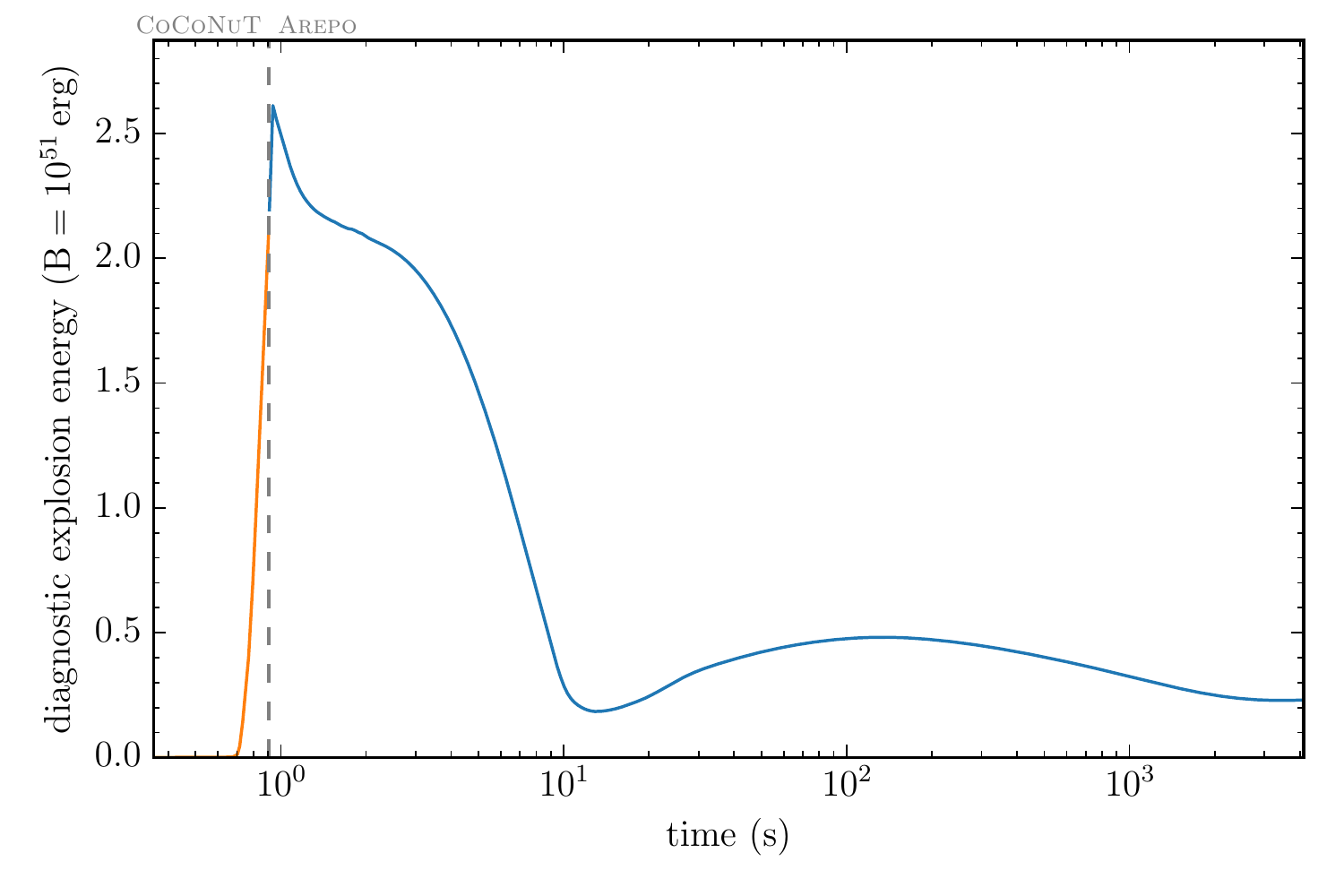}
    \caption{Diagnostic explosion energy during the \textsc{CoCoNuT} (orange) and \textsc{Arepo} (blue)  simulation.  A small discrepancy arises due to the differences in EoS, and gravity treatment.
    \label{fig:ediag}}
\end{figure}

The diagnostic explosion energy $E_\mathrm{diag}$ is evaluated as in \citet{Mueller2017} in the \textsc{CoCoNuT} run, and in the subsequent \textsc{Arepo} run the Newtonian limit \citep{Buras2006} is used,
\begin{equation}
E_\mathrm{diag}=
\int\displaylimits_{e_\mathrm{tot} > 0}
\rho e_\mathrm{tot} \, \ud V,
\end{equation}
\begin{equation}
e_\mathrm{tot}= \epsilon_\mathrm{int}+\frac{v^2}{2}-
\frac{Gm}{r},
\end{equation}
where $\rho$, $\epsilon_\mathrm{int}$, $e_\mathrm{tot}$, and $v$ are the density, internal energy density (excluding rest-mass contributions), total net energy density, and velocity, and $m$ is the enclosed mass at radius $r$. Although $E_\mathrm{diag}$ grows rapidly (Figure~\ref{fig:ediag}), BH formation due to ongoing accretion ensues $0.55\,\mathrm{s}$ after bounce.

At this time, two large neutrino-heated bubbles that expand in roughly opposite directions have pushed the shock to about $4,000\,\mathrm{km}$ (Figure~\ref{fig:initial}), and $E_\mathrm{diag}$ has reached a value of $2.09\times10^{51}\,\mathrm{erg}$.  This barely equals the binding energy, $E_\mathrm{bind}=2.1\times 10^{51}\, \mathrm{erg}$ of the shells outside the shock.  One might therefore expect that the incipient explosion is quenched after BH formation.

\begin{figure*}
\centering
	\includegraphics[width=0.75\linewidth]{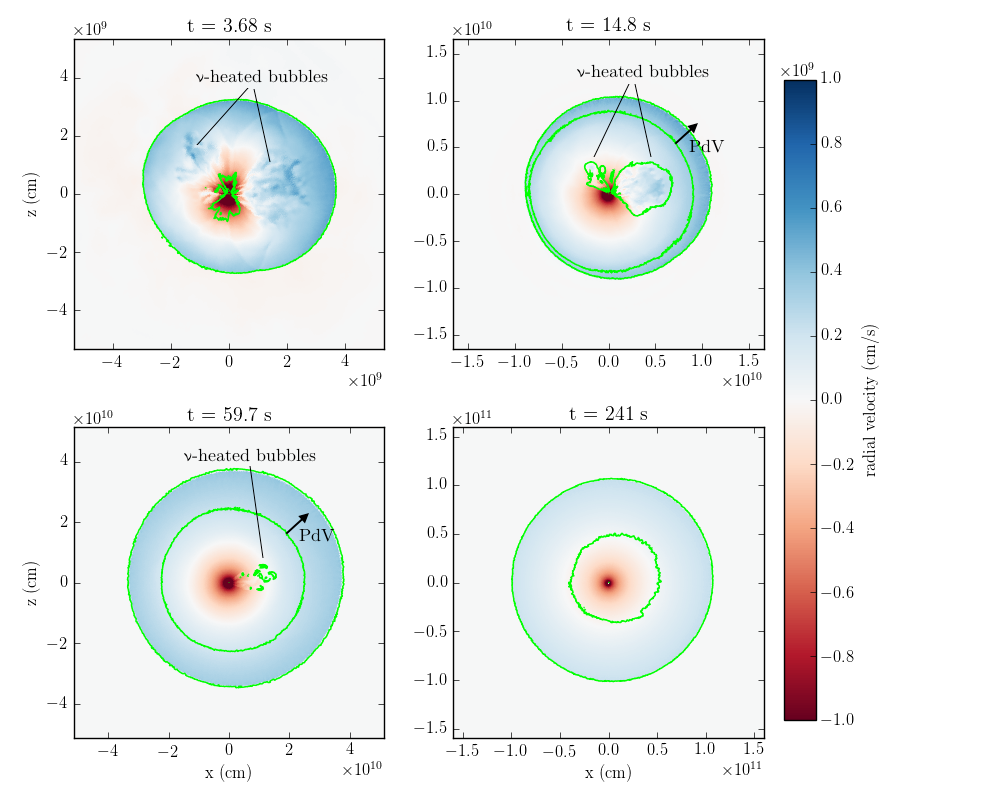}
    \caption{Radial velocity in a slice through the origin.  Green isocontours indicate zero net energy.  There exists a shell of material with negative diagnostic explosion energy that is still expanding outwards, depositing energy into the shocked material by $P\,\ud V$ work.	\label{fig:ediag_isocontours}}
\end{figure*}

\subsection{Explosion Dynamics after Black Hole Formation} 
The neutrino-heated bubbles continue to expand for a significant time after BH formation in the \textsc{Arepo} run, however, (Figure~\ref{fig:ediag_isocontours}).  Buoyancy allows these under-dense bubbles to expand against the drag of the infalling shells and maintain a positive net energy out to a radius of several $10,000\,\mathrm{km}$, although turbulent mixing and drag slowly drain the positive diagnostic energy of the bubbles during the first seconds of the explosion.  The bubbles only fall back onto the BH after tens of seconds.  This provides them with sufficient time to transfer considerable energy and momentum to the matter swept up by the shock, which develops outward velocity and positive net energy outside a radius of $\mathord{\sim}100,000\,\mathrm{km}$.  When the bubbles undergo fallback, the shock wave remains sufficiently energetic to propagate to the stellar surface with a final explosion energy of $2.3\times10^{50}\,\mathrm{erg}$.  The evolution of $E_\mathrm{diag}$ is not monotonic.  After a minimum at $\sim10\, \mathrm{s}$, it rises until $100\,\mathrm{s}$ after bounce.  This behavior is explained by Figure~\ref{fig:ediag_isocontours}: Energy can be transferred to the front of the shock wave by $P\,\mathrm{d}V$-work even from inner shells with negative net energy, as long as they still expand.  The energy transfer from the initial neutrino-heated ejecta, which eventually undergo fallback, to the propagating shock wave is operating until late times when the neutrino-heated bubbles no longer have positive net energy.

\begin{figure}
	\includegraphics[width=\columnwidth]{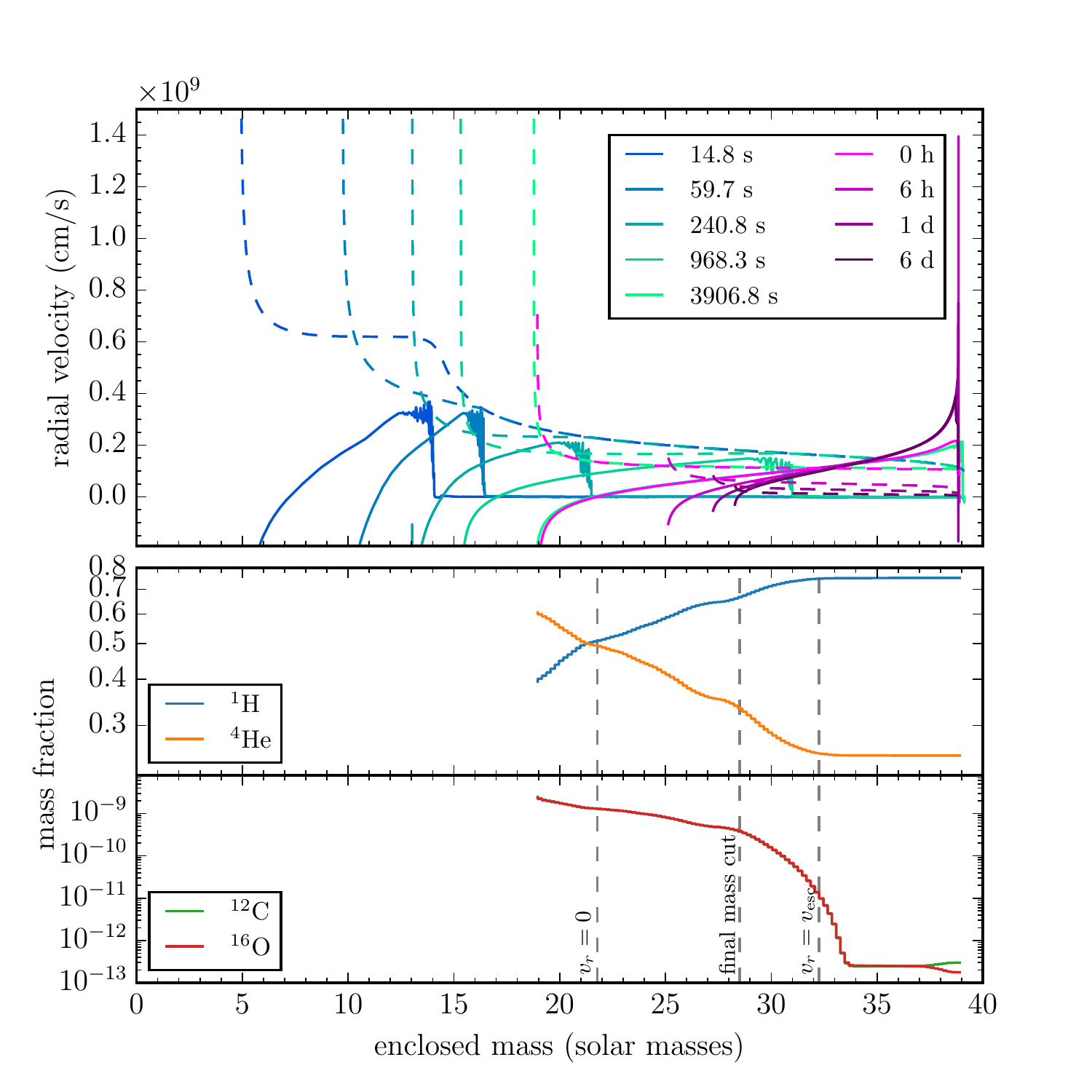}
    \caption{\textsl{Top:} Radial velocity (solid lines) and escape velocity (dashed) as a function of enclosed mass at selected times up to shock breakout (blue/cyan) and after shock breakout (magenta/purple).  \textsl{Bottom:} Mass fractions as a function of mass coordinate at shock breakout.  The dashed lines show the radii where $v_\mathrm{r}=0$ and $v_\mathrm{r}=v_\mathrm{esc}$ at breakout (\textsl{left}/\textsl{right}) and the final mass cut (\textsl{middle}). 
    \label{fig:vr_chem}}
\end{figure}

After a few tens of seconds, the shock wave becomes spherical and, as shown by radial velocity profiles in Figure~\ref{fig:vr_chem}, decelerates as it sweeps through the outer shells of the progenitor. At shock breakout, the outermost $18.2\,\Msun$ have been accelerated to positive velocity.  To determine late-time fallback, we spherically average the model, map it into the finite-volume code \textsc{Prometheus} \citep{Fryxell1991}, and evolve it until $6\,\mathrm{d}$ after shock breakout using a moving grid of $1,200$ radial zones with outflow boundary conditions at the inner boundary. The final mass cut emerges at a mass coordinate of $28.5\,\Msun$.  This is close to the point where the radial velocity equals the Keplerian orbital velocity ($1/\sqrt{2}$ of the escape velocity) at shock breakout.  No reverse shock ever forms in this explosion.  Instead, a rarefaction wave forms as the inner ejecta are decelerated by gravity and undergo fallback.

With an ejecta mass of $\mathord{\sim}11\,\Msun$, the observable transient should appear similar to a low-energy explosion of a moderately massive progenitor except for the complete absence of a nickel-powered tail.  Relatively low ejecta masses of sub-energetic nickel-poor Type IIP supernovae \citep{Spiro2014} may thus not necessarily be an argument against a fallback origin.  Future work needs to determine whether the combinations of explosion energy, ejecta mass, and non-zero nickel mass observed in contemporary supernovae can be obtained in slightly more energetic models and with different progenitors.

There is no discernible mixing during the propagation of the shock through the envelope.  The ejected material therefore contains mostly hydrogen and helium, possibly with a tiny admixture of CNO elements and Ca that emerge from Pop III star hydrogen burning \citep{Keller2014}.  This admixture is too small to explain C, O, and Mg abundance patterns in C-enhanced metal-poor (CEMP) stars like J031300.36-670839.3 with this particular realisation of a fallback supernova.  Further simulations with more favorable initial energetics are required for ejection of substantial amounts of intermediate-mass elements.

\begin{figure}
	\includegraphics[width=\columnwidth]{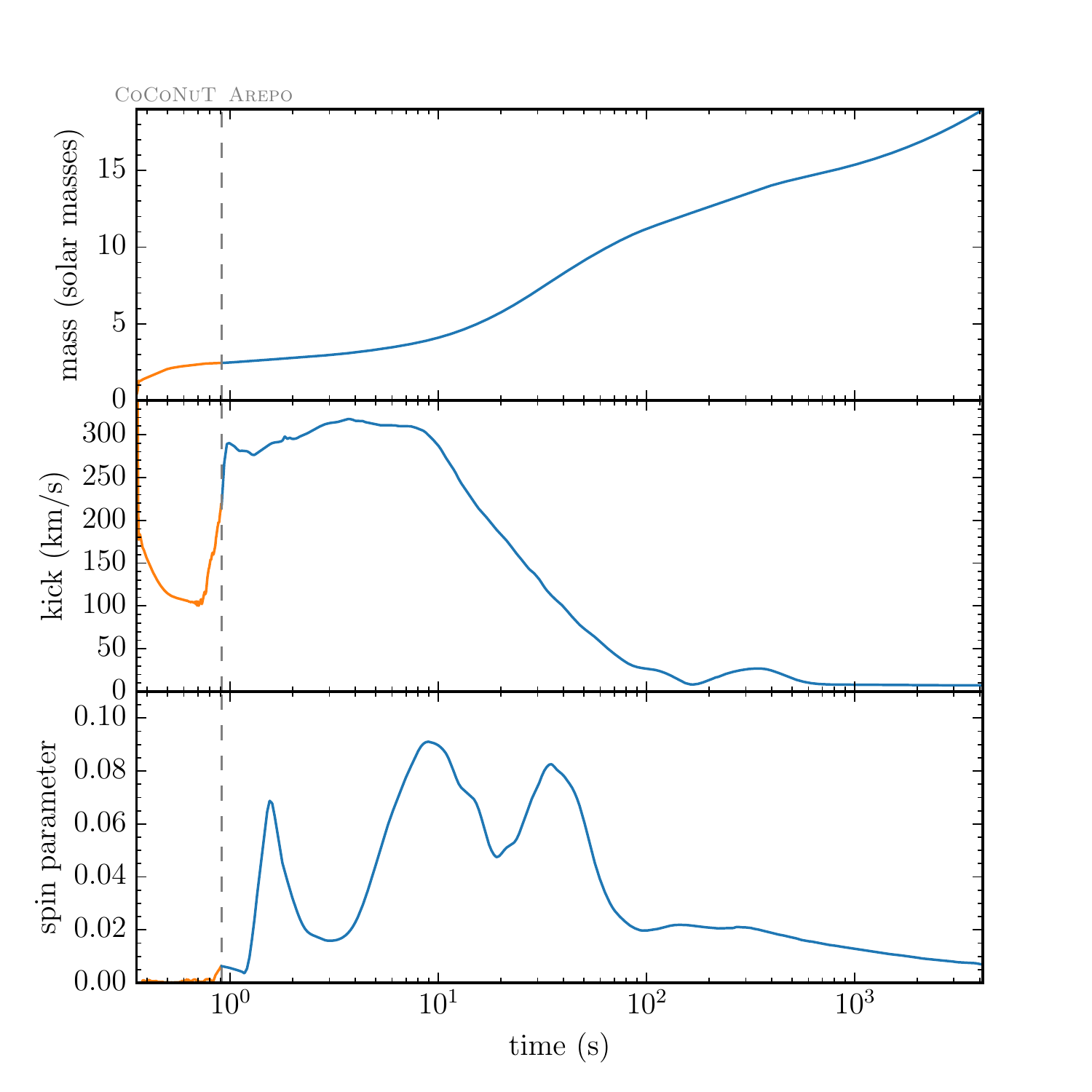}
    \caption{Top to bottom: Mass,
    kick velocity $v_\mathrm{BH}$, and spin
    parameter $a$ of the central remnant.
    \label{fig:bh_properties}}
\end{figure}

\subsection{Remnant Properties}
 
Initial asymmetries in the explosion can  contribute to the kick and spin of the BH.  We calculate the net kick onto the central remnant from the ejecta momentum assuming momentum conservation \citep{Scheck2006},
\begin{equation}
\mathbf{v}_\mathrm{BH}=
\frac1{M_\mathrm{BH}}\int_\mathrm{ejecta} \rho \mathbf{v} \,\mathrm{d}V,
\end{equation}
where $\rho$ and $\mathbf{v}$
are the density and velocity, and $M_\mathrm{BH}$ is the instantaneous mass of the BH.
The angular momentum $\mathbf{J}$ of the BH is obtained by integrating the angular momentum flux onto the central remnant:
\begin{equation}
\frac{\ud \mathbf{J}}{\ud t} = 4\pi r_*^2 \int  \rho\left(\mathbf{r}\times\mathbf{v}\right)
\ud \Omega,
\end{equation}
where $r_*$ is a fiducial radius outside the accretion radius.  For numerical evaluation this requires averaging the angular momentum flux in a thin shell around $r_*$.  From this, one can determine the BH spin parameter, $a$:  
\begin{equation}
a = \frac{cJ}{GM^2}\;.
\end{equation}

Figure~\ref{fig:bh_properties} shows $a$, BH mass, and kick velocity.  The kick is already substantial ($\mathord{\sim}200 \, \mathrm{km}\,\mathrm{s}^{-1}$) at BH formation, and continues to increase up until $10\,\mathrm{s}$.  Asymmetric accretion transiently spins up the BH to $a=0.09$.  Once the asymmetric neutrino-heated ejecta have undergone fallback and the shock wave becomes spherical, however, the BH kick and spin converge to low values.  At shock breakout we find $v_\mathrm{BH}\sim10\,\mathrm{km}\,\mathrm{s}^{-1}$ and $a \sim 0.01$. 

Our model cannot explain the high natal BH kicks of several $100\, \mathrm{km}\,\mathrm{s}^{-1}$ that have been inferred by some authors for individual black hole low-mass X-ray binaries (\citealp{RDS2012,RIN2017};  for a more critical assessment of these claims, see \citealp{Mandel2016EstimatesBinaries}).  Our results strongly suggest, however, that fallback could be a viable explanation for such large kicks if the kick were to freeze out close to the maximum values during the simulation.  This could happen if the explosion retains large asymmetries at late times, either because of incomplete fallback in a more energetic explosion, or because part of the outer shells are removed, e.g., by binary interaction. By contrast, even if the spin were to freeze out near its maximum in our model, the spin-up of the BH by fallback would be insufficient to account for the unexplained high spin parameters in some high-mass X-ray binaries
\citep{Moreno2008}.

\section{Conclusions}

We presented the first 3D simulations that follow BH formation and fallback in a core-collapse supernova from the collapse of the progenitor's iron core through shock revival by the neutrino-driven mechanism to BH formation using the neutrino hydrodynamics code \textsc{CoCoNuT-FMT} and then on to shock breakout using the moving-mesh code \textsc{Arepo}.  Our simulations, conducted for a $40\,\Msun$ Pop-III star, indicate that neutrino-driven explosions that undergo BH formation due to sustained accretion after shock revival can still shed a large fraction of the stellar envelope even if the initial explosion energy is comparable to the binding energy of the envelope.  In our case, an initial explosion energy of $2\times10^{51}\,\mathrm{erg}$ proves sufficient to eject $11\,\Msun$ with a final explosion energy of $2.3\times10^{50}\,\mathrm{erg}$.  Despite using an approximate transport scheme and a reduced neutrino-nucleon scattering rate, our model suggests that the neutrino-driven engine can become rather powerful close to BH formation, energizing explosions with powers of $\mathord{\sim}5\times10^{51}\,\mathrm{erg}\,\mathrm{s}^{-1}$ due to high accretion rates and heating efficiencies.

Our simulation of the explosion dynamics and the evolution of the remnant properties after BH formation showed that neutrino-heated ejecta can evade fallback onto the BH for several seconds, giving them sufficient time to transfer part of their energy and momentum to the outer layers behind the propagating shock wave.  In our model, few traces of the initial asymmetries survive until late stages, however. There is complete fallback of the entire He core, and none of the products from the inner shells are mixed into the ejecta region.  The only change to the primordial composition of the progenitor is a tiny enrichment in C, N, O, and Ca from hydrogen burning in Pop III stars.  This by itself would not be sufficient to explain abundances in CEMP stars such as J031300.36-670839.3.  Similarly, we do not yet find the large BH kicks that have been claimed by some authors \citep{RDS2012}.  Since the kick in our simulations briefly reaches more than $300\,\mathrm{km}\,\mathrm{s}^{-1}$, however, in principle \emph{our model strongly suggests that such high kicks are attainable by fallback.}

Our study constitutes a first step towards the exploration of BH formation and fallback based on 3D supernova explosion models.  Open questions for future investigation include whether the neutrino-driven mechanism can indeed operate in massive stars of $\mathord{\sim}40\,\Msun$.  This has to be explored using simulations with more rigorous neutrino transport and without artificial changes to the neutrino opacities.  The parameter space for fallback to produce viable models for the putative polluters of some CEMP stars, nickel-poor transients, and BHs with high kicks and spins, must be determined.  Whether fallback supernovae can explain all these phenomena
in cases with an adequate ratio between the initial explosion energy
and the binding energy of the envelope will be studied in future simulations.

\section*{Acknowledgements}

This research was undertaken with the assistance of resources and services from the National Computational Infrastructure (NCI), which is supported by the Australian Government.  It was supported by resources provided by the Pawsey Supercomputing Centre with funding from the Australian Government and the Government of Western Australia.  AH and BM were supported by ARC Future Fellowships FT120100363 (AH) and FT120100363 (BM). This work was supported by JINA-CEE through US NSF grant PHY-1430152. BM was supported by STFC grant ST/P000312/1. RP and VS acknowledge support by the European Research Council through ERC-StG grant EXAGAL-308037 and thank the Klaus Tschira Foundation. 




\bibliographystyle{yahapj}
\bibliography{ms}


\label{lastpage}
\end{document}